\documentclass[reprint,superscriptaddress,nofootinbib,showkeys,amsmath,amssymb,floatfix]{revtex4-2}
\usepackage[a4paper,text={178mm,236mm},centering]{geometry}
\usepackage{times,graphicx,siunitx}
\usepackage[colorlinks=true,linkcolor=blue,urlcolor=blue,citecolor=blue]{hyperref}

\usepackage{subfigure,dcolumn}
\usepackage{placeins}
\usepackage[T1]{fontenc}
\usepackage[english]{babel}
\makeatletter
\g@addto@macro\UrlBreaks{\do\-}
\AtBeginDocument{%
  \def\p@section{}%
  \def\p@subsection{}%
  \def\p@subsubsection{}%
  \def\@seccntformat#1{\csname the#1\endcsname\quad}%
}
\makeatother

\graphicspath{{./}}

\makeatletter
\newenvironment{widefigurehere}{%
  \par\onecolumngrid
  \begin{center}\begin{minipage}{129mm}
  \def\@captype{figure}\centering
}{%
  \end{minipage}\end{center}
  \twocolumngrid
}
\newenvironment{columnfigurehere}{%
  \par\addvspace{\intextsep}\noindent
  \begin{minipage}{\columnwidth}
  \def\@captype{figure}\centering
}{%
  \end{minipage}\par\addvspace{\intextsep}
}
\makeatother

\newcommand{\institution}[1]{%
  \ifcase#1\relax
  \or
     Institute of High Energy Physics, Chinese Academy of Sciences, Beijing 100049, China.%
  \or
    Spallation Neutron Source Science Center, Dongguan, Guangdong 523803, China.%
  \or
    Sun Yat-sen University, Guangzhou 510275, China.%
  \or
    Lanzhou University, Lanzhou 730000, China.%
  \or
    University of Chinese Academy of Sciences, Beijing 100049, China.%
  \fi
}

\begin{document}
\raggedbottom

\title{Scintillator-Based Event-Mode Neutron Imaging with Timepix4 Readout}

\author{Wenqin Yang}
\affiliation{\institution{1}}
\affiliation{\institution{2}}

\author{Xingfen Jiang}
\email[Corresponding author, ]{jiangxf@ihep.ac.cn}
\affiliation{\institution{1}}
\affiliation{\institution{2}}
\author{Junan Geng}
\affiliation{\institution{1}}
\affiliation{\institution{2}}
\affiliation{\institution{3}}

\author{Yang Liu}
\affiliation{\institution{1}}
\affiliation{\institution{2}}
\affiliation{\institution{4}}

\author{Qicai Li}
\affiliation{\institution{1}}
\affiliation{\institution{2}}
\affiliation{\institution{5}}

\author{Hongbin Liu}
\affiliation{\institution{1}}
\affiliation{\institution{2}}
\affiliation{\institution{5}}

\author{Xiaojuan Zhou}
\affiliation{\institution{1}}
\affiliation{\institution{2}}

\author{Yanfeng Wang}
\affiliation{\institution{1}}
\affiliation{\institution{2}}

\author{Li Yu}
\affiliation{\institution{1}}
\affiliation{\institution{2}}

\author{Xiuku Wang}
\affiliation{\institution{1}}
\affiliation{\institution{2}}

\author{Peixun Shen}
\affiliation{\institution{1}}
\affiliation{\institution{2}}

\author{Jiayi Ren}
\affiliation{\institution{1}}
\affiliation{\institution{2}}

\author{Yubin Zhao}
\affiliation{\institution{1}}
\affiliation{\institution{2}}
\affiliation{\institution{5}}

\author{Jianrong Zhou}
\email[Corresponding author, ]{zhoujr@ihep.ac.cn}
\affiliation{\institution{1}}
\affiliation{\institution{2}}
\affiliation{\institution{5}}

\author{Zhijia Sun}
\affiliation{\institution{1}}
\affiliation{\institution{2}}
\affiliation{\institution{5}}

\author{Yuanbo Chen}
\affiliation{\institution{1}}
\affiliation{\institution{2}}
\affiliation{\institution{5}}

\begin{abstract}
Scintillator-based event-mode neutron imaging is emerging as a powerful approach to neutron radiography. Unlike conventional integrating-mode neutron imaging, which accumulates scintillation light during an exposure, event reconstruction estimates neutron-event positions and arrival times from time-stamped optical responses. Here, this imaging approach is implemented using a self-developed Timepix4-based optical readout system and investigated through image-formation analysis and beam experiments at the China Spallation Neutron Source under different scintillator and optical configurations. The analysis explains how event localization can reduce image blur associated with the spatial extent of individual scintillation responses. Under the same 0.35\ensuremath{\times} optical configuration, event reconstruction improved the spatial resolution from 164.2 to 91.9 \ensuremath{\mu}m at 10\% modulation transfer function (MTF), while a 2.1\ensuremath{\times} configuration with a Gd\textsubscript{2}O\textsubscript{2}S:Tb scintillator achieved 26.1 \ensuremath{\mu}m at the same MTF threshold. Following event reconstruction and rejection of single-photon reconstructed events, the mean peak-normalized spectral response in the 6--8 \AA{} out-of-band interval decreased by 85.4\% relative to the raw-pixel wavelength-count spectrum, indicating reduced afterglow-related spectral background. Under the pulsed-beam conditions tested in this study, the maximum instantaneous pixel-hit rate per pixel was approximately four times its time-averaged value, providing an experimental basis for assessing readout requirements at higher neutron flux. Conventional radiography resolved structural details, while energy-selective imaging enhanced the contrast between materials with similar broadband transmission and Bragg-edge imaging revealed crystallographic features in steel. Together, the image-formation analysis and beam experiments demonstrate the imaging-quality benefits of scintillator-based event-mode neutron imaging, while sample measurements illustrate its imaging capabilities.
\end{abstract}

\keywords{Neutron imaging, Scintillators, Timepix4, Event reconstruction, Time-of-flight}

\maketitle

\section{Introduction}\label{sec:introduction}

Neutron imaging enables nondestructive visualization of internal sample structures and provides distinctive contrast for materials and engineering studies because neutron interactions vary among elements and isotopes \cite{gao2023battery,ref3,yang2020hydrogen,ref1}. Depending on the application, higher spatial resolution helps resolve structural details \cite{ref8,ref17,csnsMicron}, while time-resolved imaging follows dynamic processes \cite{ref15,carrRealtime}. Wavelength-selected images show spatial contrast within selected wavelength bands \cite{kockelmann2007,tremsin2016weld,ref16}, and transmission spectra can reveal wavelength-dependent features such as Bragg edges \cite{ref45,ref46,ref47}. Research reactors and spallation neutron sources, including the China Spallation Neutron Source (CSNS), together with dedicated imaging instruments at these facilities, provide the experimental conditions for such measurements \cite{ref23,csnsErni,cmrrFacility}. Making full use of these experimental conditions depends not only on beam performance but also on the detector's spatial localization capability \cite{wang2023mcp,parker2013}. For time-of-flight (TOF) measurements, neutron arrival times must also be recorded to obtain wavelength-dependent information.

Conventional scintillator-based neutron imaging systems typically use a mirror and imaging optics to transfer the light emitted by a scintillator screen to a CCD or CMOS camera, which accumulates scintillation light during an exposure to form an integrating-mode image \cite{carrStation,ref30,gosJiang}. By selecting the screen size and optical magnification, these systems can cover larger samples or image local regions in greater detail \cite{csnsGppd}. The transport of secondary particles and the propagation of scintillation light give the response of a single neutron interaction a finite spatial extent, which can degrade the spatial resolution of integrating-mode images \cite{baechler2002}. Reducing the scintillator-screen thickness helps limit this spatial spread but can also reduce neutron absorption and light output, motivating research into neutron absorption in thin layers and the luminescence and optical properties of scintillator materials \cite{yasuda2012}. For gadolinium oxysulfide (GOS), used in high-spatial-resolution neutron imaging, \textsuperscript{157}Gd enrichment increases neutron absorption in thin layers \cite{trtik157}, while dense ceramics have demonstrated higher light output than powder forms \cite{gosPan}. Dopant, fabrication, and processing studies have improved the luminescence and optical properties of the material \cite{gosHuang,gosWu2023,gosWu2025,tang2023gos}, and substrate engineering has also enhanced the light output of thin GOS screens \cite{ref40}. Alongside improvements to scintillator screens and optical systems, scintillator-based event-mode neutron imaging offers another route to improving spatial resolution by changing how images are formed: it identifies the light response from an individual neutron interaction and estimates its position, forming an image from reconstructed neutron-event counts rather than accumulated light intensity. Frame-based studies have achieved better spatial resolution than direct integration by acquiring scintillation light frame by frame and reconstructing neutron-capture event positions \cite{ref18,gustschin2024,ma2025}. However, these studies did not separately record the arrival times of individual neutron events within each exposure window. Combining event reconstruction with time-stamping optical readout allows scintillator-based event-mode neutron imaging to improve spatial resolution through event localization while also recovering neutron-event arrival times.

Our team previously coupled a scintillator screen to TPX3Cam and realized scintillator-based event-mode energy-resolved neutron imaging at CSNS \cite{ref33,ref35}, while other groups have also investigated related approaches \cite{ref37,ref36}. The CSNS Phase II upgrade aims to increase proton beam power on the spallation target from the Phase I design value of 100 kW to 500 kW, introducing new requirements for event readout at higher neutron flux \cite{ref11}. To address these requirements, our team developed readout electronics \cite{ref39} and the CTPX1 camera \cite{ref49} based on the Timepix4 chip \cite{ref38}, providing a self-developed readout platform for this study. In scintillator-based imaging, scintillation-light responses carry neutron spatial and temporal information, but their spatial extent and delayed emission also affect images and wavelength-count spectra formed by directly accumulating pixel hits. With the scintillator screen and optical configuration held unchanged, the present study therefore examines the extent to which forming images and spectra from reconstructed neutron events, rather than directly accumulating pixel hits, reduces the effects of spatially extended scintillation-light responses and delayed emission on the resulting imaging quality.

In this work, scintillator-based event-mode neutron imaging with Timepix4 readout is implemented and investigated using the CTPX1 camera. Image-formation analysis distinguishes the spatial extent of scintillation-light responses from the localization errors of reconstructed neutron events, providing a basis for interpreting the spatial-resolution improvement measured under the same scintillator and optical configuration. Experimental comparisons of wavelength-count spectra further show that event reconstruction and rejection of single-photon reconstructed events reduce the relative out-of-band spectral response after peak normalization, providing quantitative evidence of reduced afterglow-related spectral background. Together, these results demonstrate measurable imaging-quality benefits of forming images and spectra from reconstructed neutron events. Conventional transmission radiographs, wavelength-selected images, and Bragg-edge imaging results further demonstrate the imaging capabilities enabled by reconstructed neutron-event positions and arrival times. Measurements of the instantaneous pixel-hit load under pulsed-beam irradiation provide a basis for assessing readout requirements as this imaging approach is extended to higher neutron flux.

\section{Materials and methods}\label{sec:materials-and-methods}

\subsection{Imaging principle}\label{sec:imaging-principle}

\begin{figure}[!htb]
\centering
\includegraphics[width=\columnwidth]{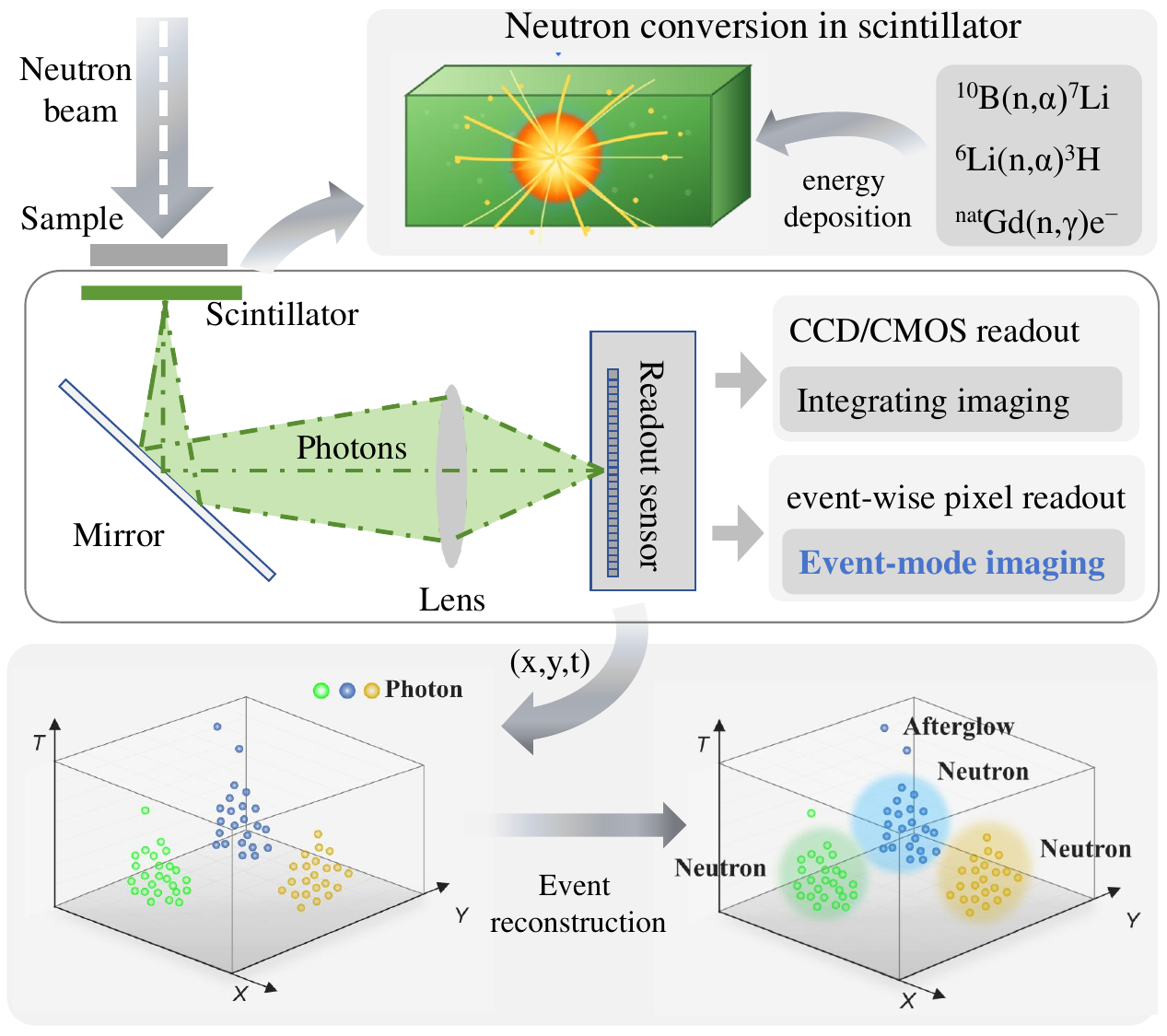}
\caption{Principle of scintillator-based event-mode neutron imaging.}
\label{fig:1}
\end{figure}

In scintillator-based neutron imaging, neutron absorption in the scintillator screen produces secondary charged particles or conversion electrons that deposit energy and excite scintillation, giving rise to a light response distributed in space and time. Conventional integrating-mode neutron imaging typically uses CCD or CMOS cameras to accumulate scintillation-light intensity during an exposure to form an image. In contrast, scintillator-based event-mode neutron imaging records pixel-level responses with spatial and timing information. Event reconstruction uses these responses to estimate the position and arrival time of individual neutron interactions, and images are formed from the reconstructed neutron events. Figure~\ref{fig:1} illustrates the principle of scintillator-based event-mode neutron imaging, including scintillation-light generation, optical transport and readout, and neutron-event reconstruction.

Let $(x_k,y_k,t_k)$ denote the true position and time of the $k$th neutron interaction, and let $L_k(x,y,t)$ denote its detectable optical response. Here, $x$, $y$, and $t$ specify where and when the response is observed. All spatial coordinates are expressed in a common imaging coordinate system. Within the linear-response range of the system, considering only the optical responses generated by neutron interactions, the light-signal component of a conventional integrating-mode neutron image over the exposure window $[t_a,t_b)$ can be expressed as
\begin{equation}
\label{eq:integrating-image}
\begin{aligned}
I_{\mathrm{int}}(x,y;t_a,t_b)
&=\int_{t_a}^{t_b}L(x,y,t)\,\mathrm{d}t\\
&=\sum_k\int_{t_a}^{t_b}L_k(x,y,t)\,\mathrm{d}t.
\end{aligned}
\end{equation}
Here, $L(x,y,t)$ is the sum of the optical responses from individual neutron interactions. Equation~\eqref{eq:integrating-image} shows that conventional integrating-mode neutron imaging accumulates the scintillation light received during the exposure window to form a spatial distribution of the light signal, but cannot separately record the arrival times of individual neutron interactions within that window.

In contrast, scintillator-based event-mode neutron imaging retains the positions and times of pixel responses. Because one neutron interaction produces multiple scintillation photons, its signal typically comprises a group of correlated responses within a finite spatial and temporal range and cannot be equated directly with a single pixel hit. Event reconstruction is therefore required to estimate the neutron-interaction position and arrival time from these responses. For an identifiable and separable single-interaction response, this process can be represented at the principle level as
\begin{equation}
\label{eq:event-reconstruction-principle}
(\hat{x}_k,\hat{y}_k,\hat{t}_k)
=\mathcal{R}\!\left[L_k(x,y,t)\right].
\end{equation}
Here, $\mathcal{R}$ is the event-reconstruction operator, and hats denote estimates of the neutron-interaction position and arrival time. Equation~\eqref{eq:event-reconstruction-principle} summarizes the key conversion from an extended optical response to a neutron event in scintillator-based event-mode imaging. In practice, reconstruction operates on the recorded discrete pixel data. The detector implementation and event-reconstruction method are described in the following sections.

Let $\mathcal{A}$ denote the set of accepted reconstructed neutron events. Their joint spatial--temporal count distribution can be expressed as
\begin{equation}
\label{eq:event-distribution}
I_{\mathrm{evt}}(x,y,t)
=\sum_{k\in\mathcal{A}}
\delta(x-\hat{x}_k)\delta(y-\hat{y}_k)\delta(t-\hat{t}_k).
\end{equation}
Here, $\delta$ is the Dirac delta distribution. Unlike integrating-mode imaging, which accumulates extended optical intensity, Eq.~\eqref{eq:event-distribution} assigns one count to the reconstructed position and time of each accepted neutron event, retaining event-wise spatial and temporal information. Actual images and spectra are obtained using finite spatial bins and time windows. Integration over a selected time window gives
\begin{equation}
\label{eq:event-window-image}
\begin{aligned}
I_{\mathrm{evt}}(x,y;t_a,t_b)
&=\int_{t_a}^{t_b}I_{\mathrm{evt}}(x,y,t)\,\mathrm{d}t\\
&=\sum_{\substack{k\in\mathcal{A}\\t_a\leq\hat{t}_k<t_b}}
\delta(x-\hat{x}_k)\delta(y-\hat{y}_k).
\end{aligned}
\end{equation}
Equations~\eqref{eq:integrating-image} and \eqref{eq:event-window-image} describe integrating-mode and event-mode neutron imaging, respectively: the former accumulates the extended optical responses generated by neutron interactions, whereas the latter accumulates neutron-event counts at reconstructed positions. When the time window covers the full valid acquisition interval, Eq.~\eqref{eq:event-window-image} yields a time-integrated neutron-event image; restricting the window yields the corresponding count image for the selected time interval.

Event reconstruction can also improve the spatial resolution of neutron images. Consider weighted-centroid localization for a correctly associated neutron event, with response-unit positions $x_{ki}$, nonnegative weights $w_{ki}$, and total weight $W_k=\sum_iw_{ki}>0$. The reconstructed position is $\hat{x}_k=(\sum_iw_{ki}x_{ki})/W_k$. By the centroid definition, the weighted displacements of the response units from the centroid sum to zero. The cross term therefore vanishes when expanding the weighted mean squared displacement about the true interaction position. Using the nonnegativity of the internal variance about the centroid gives
\begin{align}
(\hat{x}_k-x_k)^2
&=\frac{\sum_iw_{ki}(x_{ki}-x_k)^2}{W_k}
-\frac{\sum_iw_{ki}(x_{ki}-\hat{x}_k)^2}{W_k}\notag\\[4pt]
&\leq\frac{\sum_iw_{ki}(x_{ki}-x_k)^2}{W_k}.\label{eq:centroid-spatial-bound}
\end{align}
Equation~\eqref{eq:centroid-spatial-bound} shows that, for the same neutron event and the same weights, when the positively weighted responses have a nonzero spatial spread along $x$, the absolute centroid-localization error is smaller than the weighted root-mean-square displacement of the extended response from the true position. Thus, compared with integrating-mode imaging, which directly accumulates extended optical responses, event-mode imaging counts at reconstructed positions, so the internal response spread is no longer accumulated directly in the image, although localization error remains. This explains a mechanism by which event reconstruction can improve spatial resolution. The same analysis applies along $y$. The improvement in spatial resolution is quantified in Section~\ref{sec:field-of-view-and-spatial-resolution-flexibility} by comparing the modulation transfer function (MTF) curves before and after event reconstruction under the same configuration.

At pulsed neutron sources, reconstructed arrival times can be converted to neutron TOF using the source-pulse timing reference and then to neutron wavelength for a known flight path. Applying the window selection in Eq.~\eqref{eq:event-window-image} to TOF yields the spatial distribution of neutron-event counts within a selected wavelength band. Counting events by wavelength within a fixed spatial region yields the corresponding wavelength-count spectrum. Combined with reference measurements and appropriate normalization, these yield wavelength-selected transmission images and regional transmission spectra for Bragg-edge analysis.

\subsection{Detector implementation}\label{sec:detector-implementation}

To implement the event-wise optical readout described above, the present detector uses Timepix4 as the core pixelated event readout. The Timepix4 chip is bump-bonded to a 300 \ensuremath{\mu}m-thick silicon pixel sensor, forming a hybrid pixel detector with a 55 \ensuremath{\mu}m pixel pitch and a 448 \ensuremath{\times} 512 pixel matrix. The chip operates in a data-driven mode, in which each pixel works independently and provides time of arrival (ToA) and time over threshold (ToT) information after threshold triggering. The ToA records the arrival time of the pixel response, whereas the ToT is related to the signal amplitude. Timepix4 provides a nominal ToA timestamp bin width of 195 ps, a four-side buttable architecture, and a specified pixel-hit-rate capability in data-driven mode of approximately 3.6 Mhits mm\textsuperscript{-}\textsuperscript{2} s\textsuperscript{-}\textsuperscript{1}, corresponding to approximately 2.5 Ghits/s across the nominal 24.6 mm \ensuremath{\times} 28.2 mm sensitive matrix \cite{ref38}. These chip-level features provide the readout basis for fine spatial sampling, per-hit timing, high-count-rate operation, and future large-area extension in scintillator-based event-mode neutron imaging.

However, the Timepix4 silicon sensor cannot directly register single optical photons with sufficient reliability. A visible photon absorbed in silicon generates only a single electron--hole pair, whereas the pixel threshold of the Timepix4 chip is typically on the order of several hundred electrons. Therefore, direct projection of weak scintillation light onto the Timepix4 sensor would not reliably trigger pixel responses at the single-photon level. To overcome this limitation, an image intensifier is introduced before the Timepix4 readout. Incident optical photons are converted into photoelectrons at the photocathode, multiplied in the microchannel plate, and reconverted into an amplified optical output at the phosphor screen. In this way, a scintillation photon detected by the image intensifier can be transformed into a sufficiently intense optical spot, producing a compact pixel cluster on the Timepix4 sensor. The image intensifier, relay lens, and Timepix4 readout therefore together implement the pixelated event-wise optical readout required by the imaging principle described in Section~\ref{sec:imaging-principle}.

\begin{figure}[!htb]
\centering
\includegraphics[width=\columnwidth]{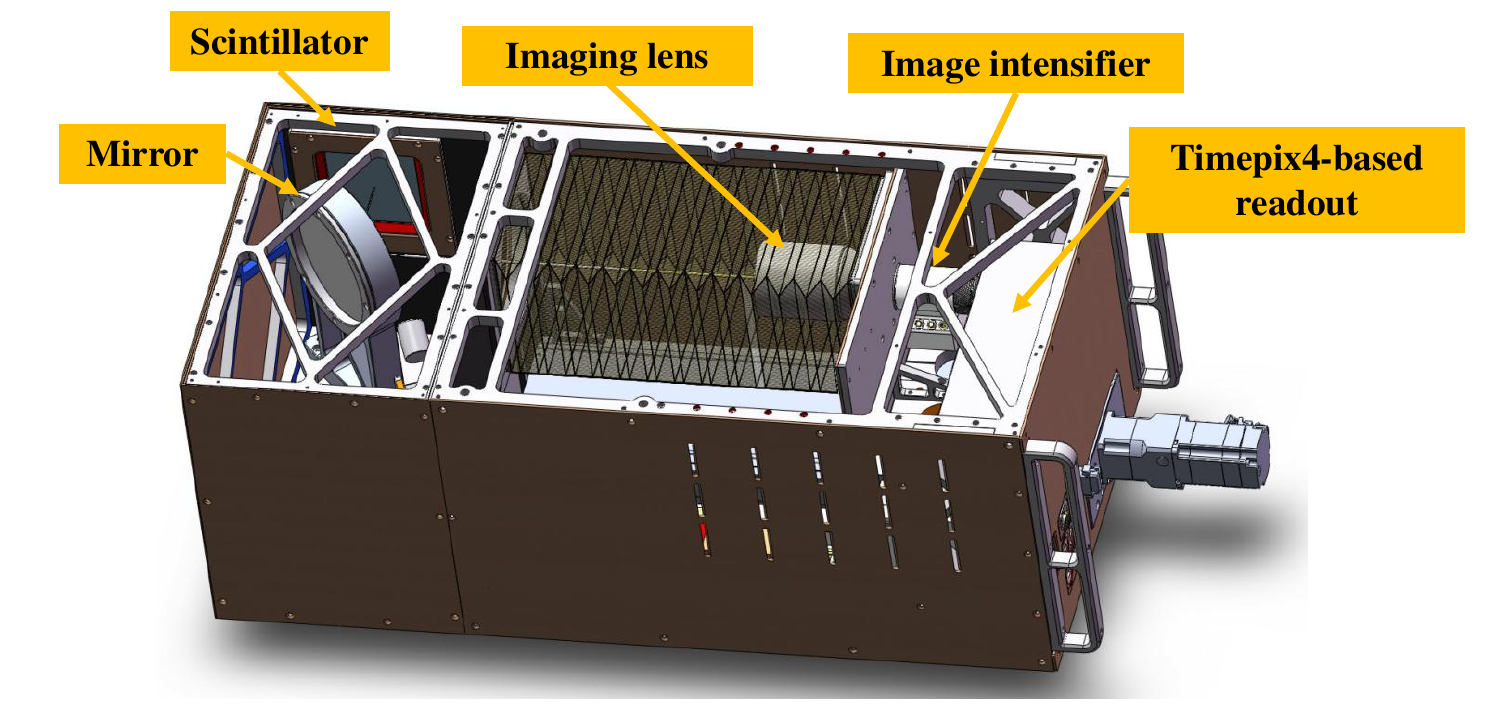}
\caption{Mechanical layout and main components of the Timepix4-based event-mode neutron imaging detector prototype.}
\label{fig:2}
\end{figure}

Based on this implementation strategy, the detector prototype was constructed using a scintillator screen, a 45\ensuremath{^{\circ}} mirror, an imaging lens, an image intensifier, a relay lens, and a Timepix4-based readout assembly, as schematically shown in Fig.~\ref{fig:2}. The optical detection chain can be divided into two sequential imaging stages. The first stage is the optical imaging from the scintillator to the photocathode of the image intensifier. Incident neutrons are first converted into scintillation photons in the neutron-sensitive screen, and the emitted photons are reflected by the 45\ensuremath{^{\circ}} mirror and focused by the imaging lens onto the photocathode. This stage maps the spatial distribution of the optical response generated by neutron interactions in the scintillator plane to the photocathode. The second stage is the optical imaging from the image intensifier to the Timepix4 sensor. The image intensifier amplifies the photon signal entering the photocathode and forms an intensified optical image on its phosphor screen, which is then relayed onto the Timepix4 sensor by the relay lens. Through these two stages, the optical signals generated by neutron interactions in the scintillator are converted into amplified pixel-level responses that can be recorded by the Timepix4 readout. These amplified pixel-level responses constitute the raw data used for the event reconstruction described in the next section.

\subsection{Event reconstruction}\label{sec:event-reconstruction}

The optical readout described above records amplified scintillation photons as pixel hits on the Timepix4 sensor. Because the image intensifier operates in a single-photon--sensitive regime, each detected photon is converted, after amplification, into a small spatial cluster on the Timepix4 sensor. Consequently, the raw pixel data simultaneously contain two hierarchical structures:

(1) single-photon pixel clusters formed by individual scintillation photons after image-intensifier amplification; and

(2) multi-photon correlations arising from multiple scintillation photons produced by a single neutron event.

\begin{figure}[!htb]
\centering
\includegraphics[width=\columnwidth]{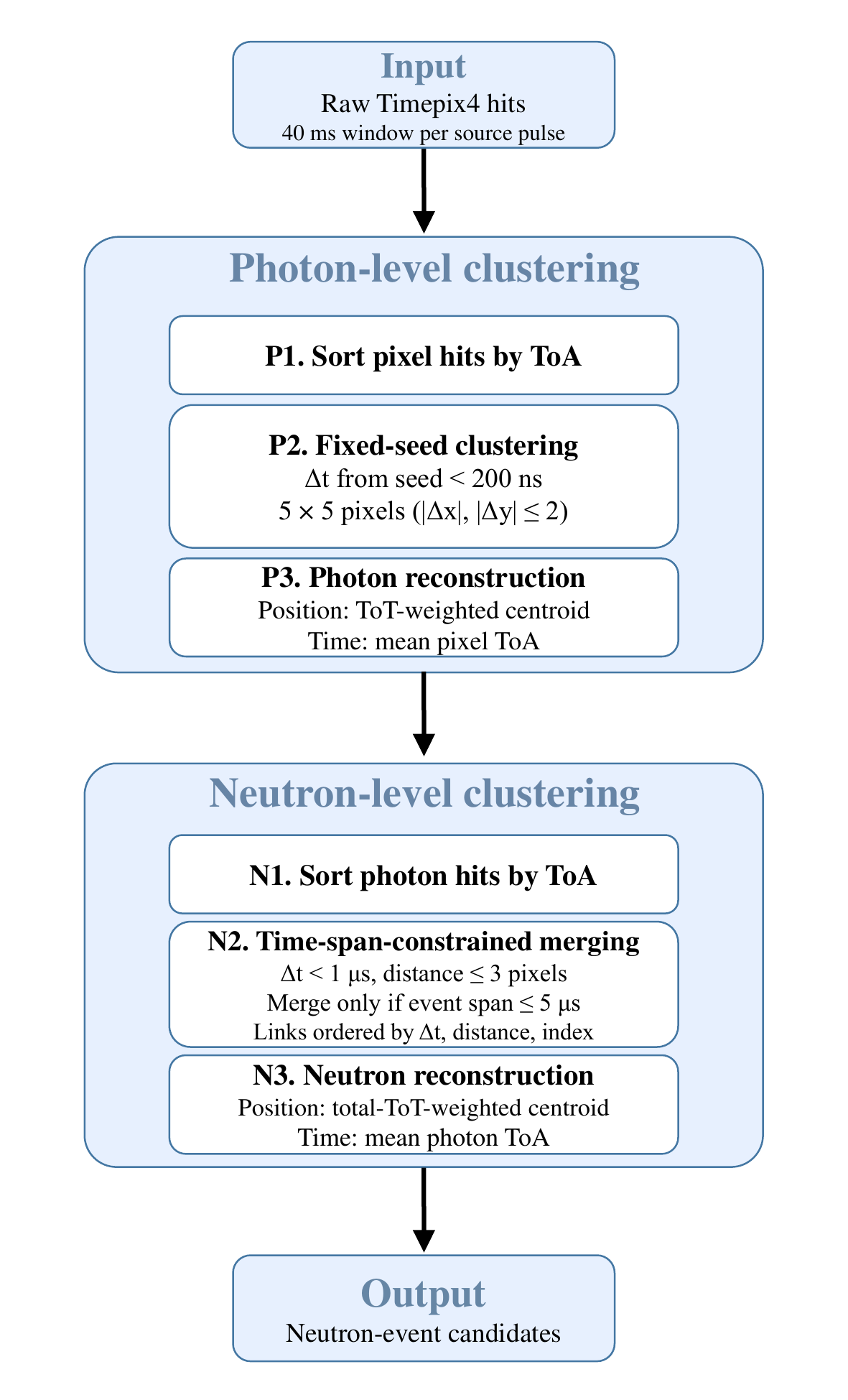}
\caption{Flowchart of the two-level photon--neutron event reconstruction method.}
\label{fig:3}
\end{figure}

To recover neutron-event information from these raw pixel data, a two-level clustering framework is used, consisting of fixed-seed photon-level clustering followed by time-span-constrained neutron-level clustering. Both stages are performed independently within each 40 ms acquisition window corresponding to a source pulse, without clustering across pulse periods. The first reconstruction step identifies individual photon hits from compact pixel clusters, and the second reconstruction step groups correlated photon hits into reconstructed neutron events. The complete processing workflow is illustrated in Fig.~\ref{fig:3}.

\subsubsection{Photon-level clustering}\label{sec:photon-level-clustering}

The purpose of photon-level clustering is to convert the compact pixel clusters produced by individual amplified scintillation photons into photon hits. All pixel hits recorded by Timepix4 are first sorted in ascending order of ToA. The clustering is then performed hit by hit. When an unassigned pixel hit is encountered, a new photon cluster is initiated, and this hit is used as the seed hit of the cluster. A single scintillation photon, after amplification by the image intensifier, produces a compact pixel footprint on the Timepix4 sensor. The pixel hits within this footprint are spatially localized and temporally correlated. Photon-level clustering uses a fixed-seed strategy: candidate pixel hits are compared directly with the seed hit using a ToA-difference threshold of 200 ns and the spatial constraints defined below, rather than added through transitive growth. Subsequent pixel hits in the ToA-sorted list are examined to determine whether they belong to the current photon cluster according to local temporal and spatial constraints.

For a candidate pixel hit with coordinates (\textit{x}\textsubscript{c}, \textit{y}\textsubscript{c}) and arrival time ToA\textsubscript{c}, and a seed hit with coordinates (\textit{x}\textsubscript{0}, \textit{y}\textsubscript{0}) and arrival time ToA\textsubscript{0}, the candidate pixel hit is assigned to the current photon cluster when the following conditions are simultaneously satisfied:

\begin{equation}
\label{eq:photon-clustering}
\left\{
\begin{aligned}
\left|\mathrm{ToA}_{c}-\mathrm{ToA}_{0}\right| &< \SI{200}{\nano\second},\\
\left|x_{c}-x_{0}\right| &\leq 2,\\
\left|y_{c}-y_{0}\right| &\leq 2.
\end{aligned}
\right.
\end{equation}

Pixel hits satisfying these conditions are merged into the current photon cluster. Because the pixel hits are sorted by ToA, pixel hits outside the temporal window are not further examined for the current seed hit. After the photon cluster is formed, the photon position is calculated as the ToT-weighted centroid of its pixel hits, and the photon ToA is taken as the average ToA of the pixel hits in the cluster. This procedure compresses the pixel-level raw responses into a list of individual photon hits.

\subsubsection{Neutron-level clustering}\label{sec:neutron-level-clustering}

After the photon hit list is obtained, neutron-level clustering groups spatially and temporally correlated photon hits into candidate neutron events. All photon hits are first sorted in ascending order of ToA and initialized as separate components. Unlike photon-level clustering, neutron-level clustering employs connected-component merging subject to a maximum event time span. This is because the scintillation photons generated by a single neutron interaction may spread over a broader and less regular spatiotemporal region than the compact pixel footprint of a single amplified photon. For two photon hits with coordinates \((x_c,y_c)\) and \((x_g,y_g)\) and arrival times \(T_c\) and \(T_g\), respectively, a candidate link is established when both temporal and spatial constraints are satisfied.

\begin{equation}
\label{eq:neutron-clustering}
\left\{
\begin{aligned}
\left|T_{c}-T_{g}\right| &< T_{w},\\
\sqrt{\left(x_{c}-x_{g}\right)^{2}+\left(y_{c}-y_{g}\right)^{2}} &\leq D_{n},
\end{aligned}
\right.
\end{equation}

where \(T_w\) is the temporal neighborhood window, and \(D_n\) is the spatial neighborhood radius. In this work, \(T_w\) was set to 1 \ensuremath{\mu}s, and \(D_n\) was set to 3.0 pixels. Unlike the integer pixel offsets used in photon-level clustering, \(D_n\) represents a continuous distance in pixel units. Candidate links are processed in ascending order of time difference, with ties resolved by spatial distance and then photon indices. The linked components are merged only if the resulting event satisfies \(T_{\max}-T_{\min}\leq5\ \mu\mathrm{s}\), where \(T_{\max}\) and \(T_{\min}\) are the latest and earliest constituent photon ToAs. Otherwise, the link is rejected. After clustering, the neutron-event position is calculated as the ToT-weighted centroid of all photon hits in the cluster, using each photon's summed pixel ToT as its weight. The neutron-event ToA is taken as the arithmetic average ToA of its constituent photon hits and is used for subsequent TOF-based wavelength or energy reconstruction.

\subsubsection{Afterglow suppression}\label{sec:afterglow-suppression}

Scintillator afterglow originates from slow decay components of the scintillation emission. Although the delayed emission is much weaker than the prompt scintillation pulse, it can still produce delayed scintillation photons. Because the image intensifier is sensitive to single photons, afterglow photons can also be amplified and recorded as photon-like pixel clusters on the Timepix4 sensor. In the two-level reconstruction framework, these afterglow photons may be identified as photon hits after photon-level clustering. However, during neutron-level clustering, they usually appear as isolated photon hits because they lack temporally and spatially correlated neighboring photon hits within the clustering window. As a result, they tend to form single-photon reconstructed events rather than multi-photon neutron events.

Single-photon reconstructed events cannot be strictly classified as afterglow events on an event-by-event basis. A genuine neutron interaction generally produces multiple scintillation photons, but only one photon may occasionally be detected and recorded because of photon statistics, optical collection efficiency, optical transmission losses, photocathode quantum efficiency, and triggering efficiency. Therefore, in this work, reconstructed events containing only one photon are normally discarded as a statistical afterglow-suppression step. This criterion also rejects genuine neutron events represented by only one detected photon. However, it preferentially removes isolated delayed-photon events and effectively reduces the contribution of scintillator afterglow to the reconstructed neutron wavelength spectrum.

\subsection{Experimental setup}\label{sec:experimental-setup}

To investigate scintillator-based event-mode neutron imaging with Timepix4 readout under different operating configurations, beamline experiments were performed using different scintillator and optical-magnification settings. The experiments were designed to evaluate this imaging approach in terms of event timing and wavelength measurement, field-of-view/spatial-resolution flexibility, pixel-hit-rate behavior, and representative imaging performance under pulsed neutron beam conditions. Three optical magnifications, 0.14\ensuremath{\times}, 0.35\ensuremath{\times}, and 2.1\ensuremath{\times}, were used to cover different imaging working points. Throughout this work, optical magnification refers to the overall magnification from the scintillator plane to the Timepix4 sensor plane, including the 1:1 relay optics. For the 0.14\ensuremath{\times} and 0.35\ensuremath{\times} configurations, a 50 \ensuremath{\mu}m-thick ZnS:Cu/\textsuperscript{6}LiF scintillator was used. For the 2.1\ensuremath{\times} high-magnification configuration, a 10 \ensuremath{\mu}m-thick Gd\textsubscript{2}O\textsubscript{2}S:Tb (GOS:Tb) scintillator was used. The combination of different optical magnifications and scintillator screens allowed the Timepix4-readout imaging approach to be investigated under large-field, intermediate, and high-spatial-resolution configurations.

Different samples and imaging tasks were arranged according to these configurations. Under the 0.14\ensuremath{\times} magnification configuration, conventional neutron radiography was performed using a cadmium plate with a CSNS logo to demonstrate the basic spatial imaging capability. Energy-selective imaging experiments were also carried out using seven metallic samples, including cobalt (Co), molybdenum (Mo), tungsten (W), iron (Fe), nickel (Ni), copper (Cu), and brass. Under the 0.35\ensuremath{\times} magnification configuration, a mechanical wristwatch was imaged to examine fine structural visualization, and an austenitic steel sample was used for Bragg-edge imaging. Under the 2.1\ensuremath{\times} magnification configuration, the thin GOS:Tb scintillator was used for high-spatial-resolution imaging. Natural gadolinium has a higher thermal neutron capture cross section than \textsuperscript{6}Li, and the secondary electrons generated in Gd-based scintillators have ranges on the order of micrometers. These characteristics enable the use of thinner scintillator screens, which helps reduce light spreading and supports high-spatial-resolution imaging under high-magnification optical coupling.

\begin{figure*}[!t]
\centering
\includegraphics[width=129mm]{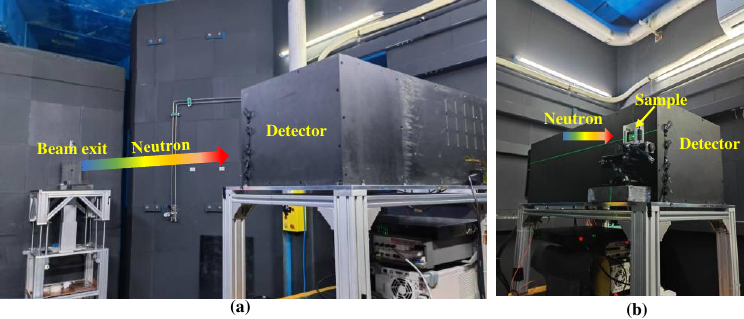}
\caption{Experimental setup of the neutron imaging system at the beamline: (a) overall layout from the beam exit to the detector; (b) configuration with the sample in the neutron beam.}
\label{fig:4}
\end{figure*}

The beamline experiments were performed at CSNS BL20. The repetition rate of the proton beam pulses striking the target is 25 Hz, corresponding to a neutron production period of 40 ms. BL20 is a test beamline designed for detector performance evaluation, detector calibration, neutron polarization studies, and neutron focusing-lens investigations. The beam exit of BL20 is located 11.4 m from the target, with a beam aperture of 36 mm \ensuremath{\times} 40 mm. A neutron chopper is installed on the beamline to select neutron wavelengths, with a fixed wavelength bandwidth of 4.48 \AA{}. In the present experiments, the detector was positioned at a total neutron flight distance of 14.956 m to obtain a larger neutron beam spot at the detector position. The selected neutron wavelength range was from 1.00 \AA{} to 5.48 \AA{}. The beamline layout and sample--detector arrangement are shown in Fig.~\ref{fig:4}.

\section{Results and discussion}\label{sec:results-and-discussion}

\subsection{Event timing and wavelength measurement}\label{sec:event-timing-and-wavelength-measurement}

\begin{figure}[!htb]
\centering
\includegraphics[width=\columnwidth]{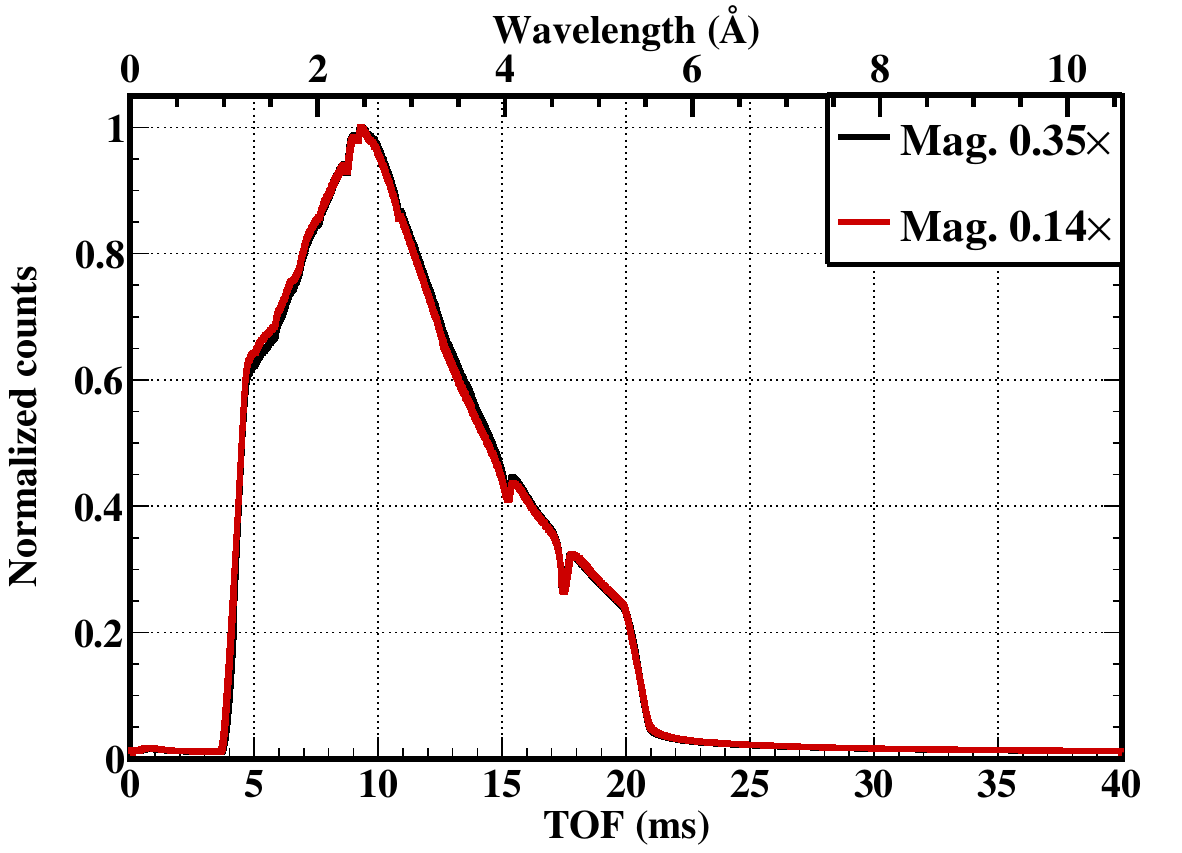}
\caption{Normalized TOF and wavelength-count spectra derived from raw pixel hits at optical magnifications of 0.14\ensuremath{\times} and 0.35\ensuremath{\times}. Each spectrum is normalized to its own maximum.}
\label{fig:5}
\end{figure}

The temporal and wavelength distributions of raw pixel hits were compared under different optical-magnification configurations. Figure~\ref{fig:5} shows the peak-normalized TOF and wavelength-count spectra obtained using the ZnS:Cu/\textsuperscript{6}LiF scintillator at magnifications of 0.14\ensuremath{\times} and 0.35\ensuremath{\times}. Because CSNS operates at a pulsed repetition rate of 25 Hz, the available TOF window spans 0--40 ms for each source period. For these raw-pixel spectra, TOF values were calculated from pixel-hit arrival times relative to the source-pulse timing reference and converted to wavelength using the neutron flight path. The obtained wavelength-count spectra are predominantly concentrated within the chopper-selected range of 1.00--5.48 \AA{}. In addition, the dips associated with aluminum Bragg edges at 4.05 \AA{} and 4.68 \AA{} are clearly observed, providing a consistency check on the TOF-to-wavelength conversion. These results show that the main wavelength-dependent features of the incident neutron beam are retained in the raw-pixel spectra under both optical configurations.

\begin{figure}[!htb]
\centering
\includegraphics[width=\columnwidth]{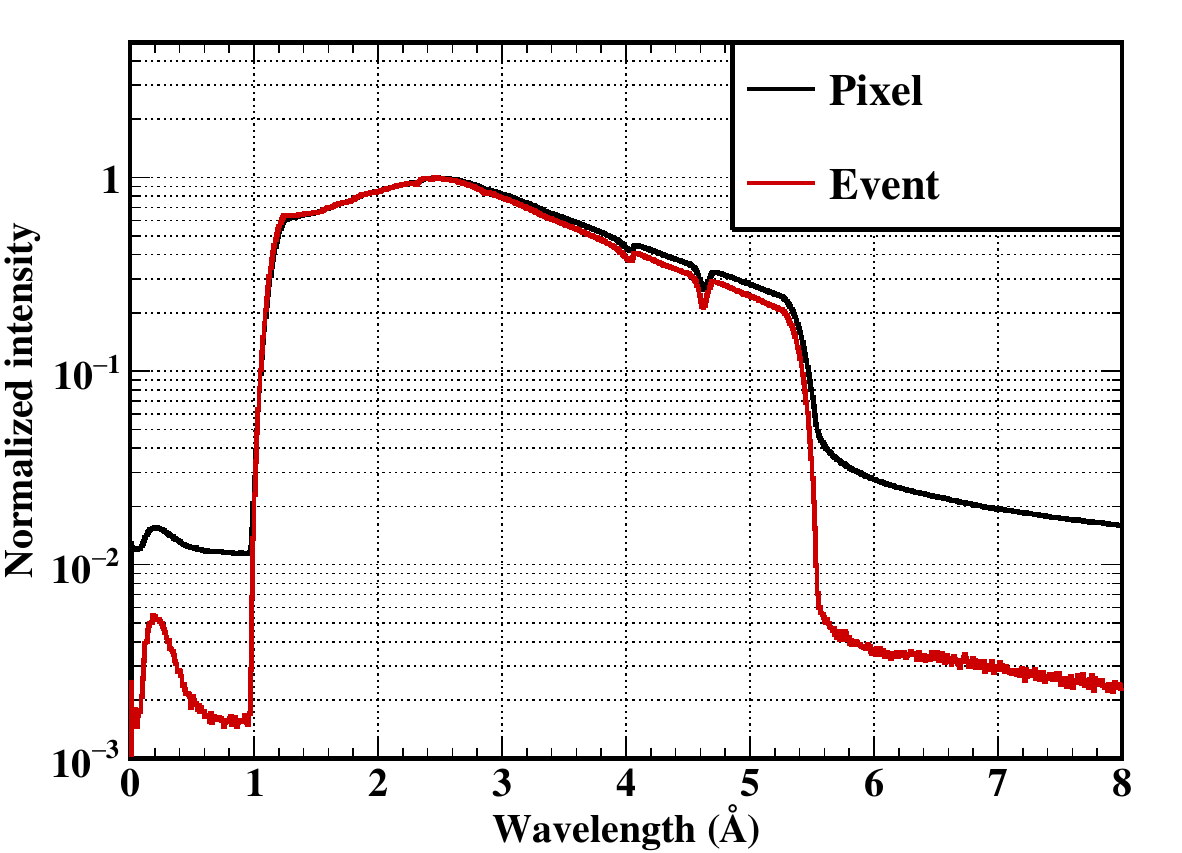}
\caption{Raw-pixel and reconstructed neutron-event wavelength-count spectra, each normalized to its own maximum.}
\label{fig:6}
\end{figure}

Under the 0.35\ensuremath{\times} magnification configuration, the raw-pixel and reconstructed neutron-event wavelength-count spectra were compared. In the reconstructed neutron-event wavelength-count spectrum, only events containing more than one photon hit were counted according to the afterglow-suppression criterion described in Section~\ref{sec:afterglow-suppression}. As shown in Fig.~\ref{fig:6}, the wavelength-count spectra normalized to their respective maxima before and after event reconstruction exhibit consistent wavelength ranges, and the positions of the aluminum Bragg-edge features remain unchanged. In contrast, the raw-pixel wavelength-count spectrum shows higher normalized counts outside the 1.00--5.48 \AA{} chopper-selected wavelength range. These excess counts are primarily attributed to delayed scintillator afterglow photons, which tend to appear as isolated photon responses. To quantify this difference, the peak-normalized spectral values were averaged over the 6--8 \AA{} out-of-band interval. Following event reconstruction and rejection of single-photon reconstructed events, the mean normalized response decreased from $2.01\times10^{-2}$ in the raw-pixel spectrum to $2.94\times10^{-3}$ in the reconstructed neutron-event spectrum, corresponding to a relative reduction of 85.4\%.

\subsection{Field-of-view and spatial-resolution flexibility}\label{sec:field-of-view-and-spatial-resolution-flexibility}

The field of view (FOV) and spatial resolution define the configurable operating range of scintillator-based neutron imaging. In the Timepix4-readout event-mode implementation, these parameters are mainly determined by the optical magnification, scintillator screen, image-intensifier active area, and event reconstruction. A Schneider lens with an f-number of 2.2 and a focal length of 50 mm was used to set the optical magnification to 0.14\ensuremath{\times}. Based on this magnification and the Timepix4 chip dimensions, the theoretical imaging field of view is 176 mm \ensuremath{\times} 201 mm. The measured large-FOV imaging result and spatial-resolution evaluation are shown in Fig.~\ref{fig:7}. The valid neutron counts are confined within a circular region with a diameter of approximately 129 mm, while the responses outside this region are likely associated with cosmic-ray background. This circular field is determined by the 18 mm active diameter of the image intensifier, which is projected onto the scintillator plane by the 0.14\ensuremath{\times} optical magnification.

In Fig.~\ref{fig:7}(a), the square region intersecting the circular field corresponds to the 100 mm \ensuremath{\times} 100 mm scintillator outline, while the actual illuminated region is further limited by the beam spot at the test position. Figs.~\ref{fig:7}(b) and \ref{fig:7}(c) show the spatial-resolution results before and after event reconstruction, respectively. Based on visual inspection, features of approximately 500 \ensuremath{\mu}m are resolved before reconstruction, whereas 300 \ensuremath{\mu}m features can be resolved after reconstruction. Considering the 0.14\ensuremath{\times} optical magnification, the effective pixel pitch at the scintillator plane before reconstruction is 393 \ensuremath{\mu}m, indicating that the raw-pixel image resolution is mainly limited by pixel sampling. After reconstruction, the ToT-weighted event centroids are mapped onto a finer 2D matrix with a bin size equal to one-third of the effective pixel pitch, corresponding to a reconstructed image bin size of 131 \ensuremath{\mu}m. These results demonstrate that event reconstruction improves spatial detail visibility even in the large-FOV configuration.

\begin{widefigurehere}
\centering
\includegraphics[width=129mm]{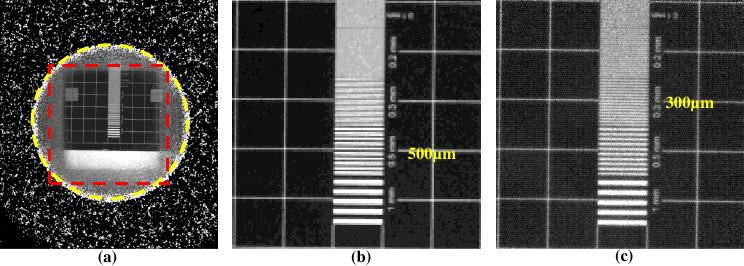}
\caption{Neutron images of the modulation transfer function (MTF) bar pattern at a magnification of 0.14\ensuremath{\times}: (a) full-field neutron image; (b) raw-pixel image of the central region; (c) corresponding reconstructed neutron-event image.}
\label{fig:7}
\end{widefigurehere}

\begin{figure}[!htb]
\centering
\includegraphics[width=\columnwidth]{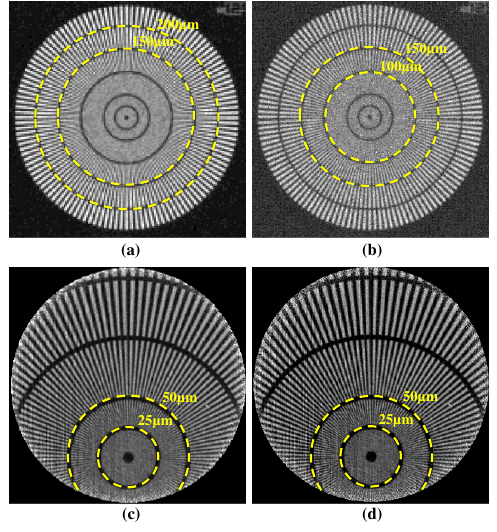}
\caption{Siemens star resolution target images at two magnification settings: (a) raw-pixel image at 0.35\ensuremath{\times}; (b) reconstructed neutron-event image at 0.35\ensuremath{\times}; (c) raw-pixel image at 2.1\ensuremath{\times} with a 10 \ensuremath{\mu}m GOS:Tb scintillator; (d) reconstructed neutron-event image at 2.1\ensuremath{\times}.}
\label{fig:8}
\end{figure}

\begin{figure}[!htb]
\centering
\includegraphics[width=\columnwidth]{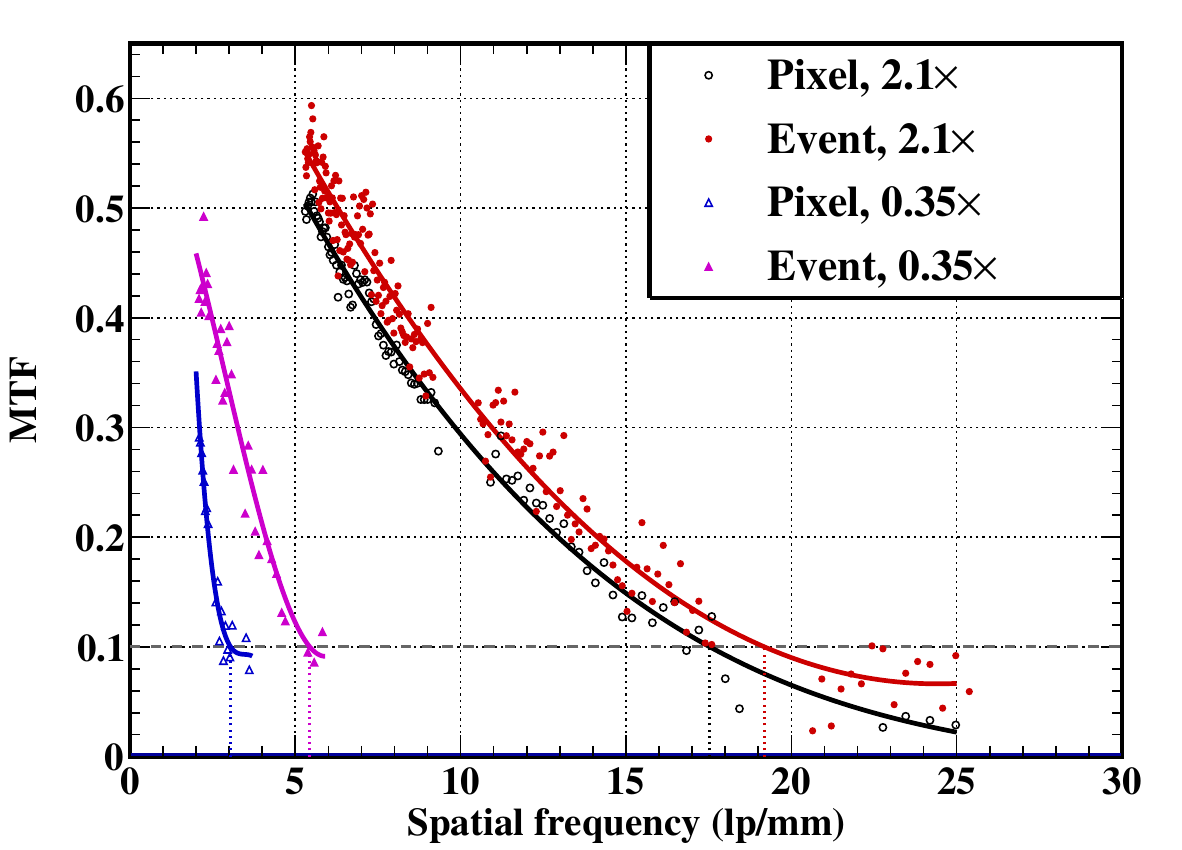}
\caption{MTF curves before and after event reconstruction at magnifications of 0.35\ensuremath{\times} and 2.1\ensuremath{\times}.}
\label{fig:9}
\end{figure}

To further investigate the spatial-resolution performance under smaller fields of view, measurements were performed at optical magnifications of 0.35\ensuremath{\times} and 2.1\ensuremath{\times}. In these configurations, a Siemens star test target was used to provide continuously varying spatial frequencies. At 0.35\ensuremath{\times} magnification, the raw-pixel and reconstructed neutron-event images are shown in Figs.~\ref{fig:8}(a) and \ref{fig:8}(b). For the high-resolution 2.1\ensuremath{\times} configuration, a 10 \ensuremath{\mu}m-thick GOS:Tb scintillator containing natural gadolinium was employed, and a 10 cm-thick boron brick with a 2 cm aperture was placed at the beamline exit to improve beam collimation. The corresponding raw-pixel and reconstructed neutron-event images are shown in Figs.~\ref{fig:8}(c) and \ref{fig:8}(d). In both magnification settings, event reconstruction enhances detail visibility and improves the discernibility of high-spatial-frequency structures. Based on the images in Fig.~\ref{fig:8}, MTF curves were calculated, as shown in Fig.~\ref{fig:9}. The spatial resolution was calculated as the reciprocal of twice the spatial frequency at 10\% MTF. At 0.35\ensuremath{\times} magnification, event reconstruction improved the spatial resolution from 164.2 \ensuremath{\mu}m to 91.9 \ensuremath{\mu}m, corresponding to 5.44 lp/mm after reconstruction. This improvement with the scintillator screen and optical configuration unchanged provides experimental evidence for the spatial imaging benefit of event reconstruction, consistent with the image-formation analysis in Section~\ref{sec:imaging-principle}. Under the 2.1\ensuremath{\times} high-magnification configuration, the reconstructed MTF curve indicates that a spatial resolution of 26.1 \ensuremath{\mu}m can be achieved (19.2 lp/mm). Together with the large-FOV results, these measurements demonstrate that scintillator-based event-mode neutron imaging retains the flexibility to select the imaging scale through optical configuration while benefiting from event localization.

The FOV and spatial resolution are inherently coupled through the optical magnification and scintillator response. In the current implementation, the low-magnification configuration provides large-FOV imaging using a 100 mm \ensuremath{\times} 100 mm scintillator, with the usable imaging region determined by the overlap of the scintillator area, the circular optical FOV, and the illuminated region. The 2.1\ensuremath{\times} high-magnification GOS:Tb configuration provides a spatial resolution of 26.1 \ensuremath{\mu}m over an imaging area of approximately 6 mm \ensuremath{\times} 6 mm. The present system is still limited by the 18 mm active diameter of the image intensifier, which does not fully match the nominal Timepix4 sensitive matrix (24.6 mm \ensuremath{\times} 28.2 mm), corresponding to a geometric area ratio of 36.7\%. A larger-diameter image intensifier matching the Timepix4 sensitive area would improve the effective use of pixels and could either provide finer sampling at the same FOV or extend the FOV while maintaining the current effective pixel size.

\subsection{Pixel-hit-rate behavior}\label{sec:count-rate-behavior}

\begin{figure}[!htb]
\centering
\includegraphics[width=\columnwidth]{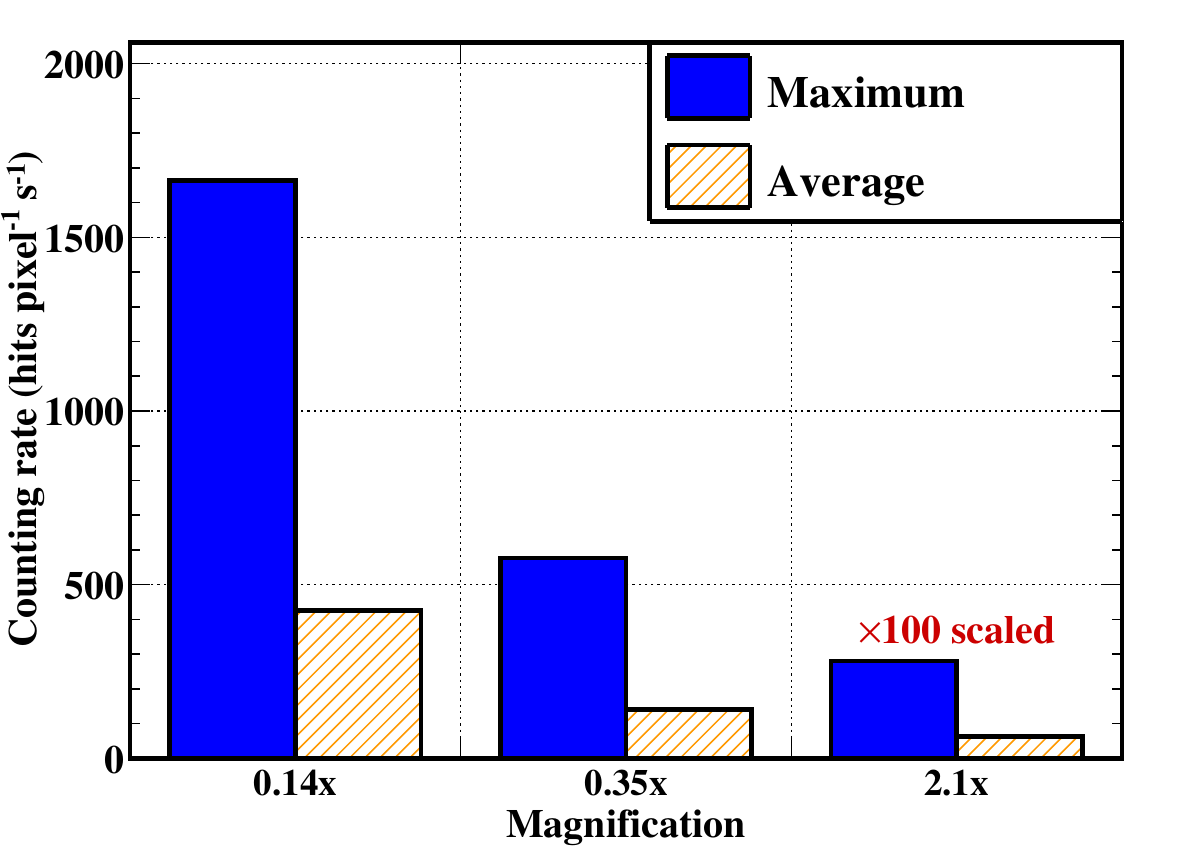}
\caption{Time-averaged and maximum instantaneous pixel-hit rates per pixel under different magnification settings. Both the time-averaged and maximum instantaneous pixel-hit rates for the 2.1\ensuremath{\times} configuration are multiplied by 100 for display.}
\label{fig:10}
\end{figure}

For event-mode neutron imaging at pulsed sources, the temporal distribution of pixel hits determines the average and peak loads that the readout system must accommodate. Here, pixel-hit rates are calculated from raw Timepix4 pixel hits, not reconstructed neutron events. For the three optical-magnification configurations described above, the time-averaged pixel-hit rate and maximum instantaneous pixel-hit rate per pixel were analyzed using data acquired without spatial-resolution test targets. For each optical configuration, raw Timepix4 pixel hits were analyzed within a 10 \ensuremath{\times} 10-pixel region of interest (ROI) at the center of the beam spot. The time-averaged pixel-hit rate per pixel was calculated by dividing the total number of pixel hits in the ROI by the effective acquisition time and the number of pixels. To determine the instantaneous rate, pixel hits from the ROI were accumulated in a TOF histogram over repeated source pulses. The count in each TOF bin was normalized by the number of effective source pulses, the bin width, and the number of pixels in the ROI. The maximum of the resulting ROI-averaged per-pixel rate distribution was taken as the maximum instantaneous pixel-hit rate per pixel. The corresponding time-averaged and maximum instantaneous pixel-hit rates are shown in Fig.~\ref{fig:10}. For a given incident neutron flux and otherwise unchanged detector conditions, lower magnification produces a higher time-averaged pixel-hit rate per pixel because each pixel corresponds to a larger effective detection area at the scintillator plane. In addition, the instantaneous pixel-hit rate is significantly higher than the time-averaged pixel-hit rate because of the intrinsic time structure of the pulsed neutron beam; in this experiment, the ratio between them is approximately 4.

The pixel-hit-rate behavior also depends strongly on the scintillator and optical configuration. GOS:Tb is commonly used for high-spatial-resolution neutron imaging, but its light output is much lower than that of ZnS:Cu/\textsuperscript{6}LiF scintillators. In the present experiment, the high-resolution GOS:Tb configuration combined higher optical magnification with additional beam collimation. The higher magnification means that each pixel corresponds to a smaller effective area at the scintillator plane. Under these operating conditions, the GOS:Tb configuration exhibits a much lower pixel-hit rate; the plotted values for this configuration in Fig.~\ref{fig:10} are multiplied by 100 for display. In contrast, the 50 \ensuremath{\mu}m-thick ZnS:Cu/\textsuperscript{6}LiF scintillator used in the 0.14\ensuremath{\times} and 0.35\ensuremath{\times} configurations provides a higher pixel-hit rate, and increasing the scintillator thickness could further enhance neutron absorption efficiency and the overall pixel-hit rate.

The measured pixel-hit-rate behavior is governed by several coupled factors, including neutron flux, optical magnification, numerical aperture of the imaging lens, scintillator material and thickness, and the TOF structure of the beamline. Higher neutron flux increases the number of incident neutrons per unit time. Lower optical magnification increases the area at the scintillator plane corresponding to each pixel, thereby increasing the per-pixel hit rate. A larger numerical aperture improves light-collection efficiency and increases the probability that scintillation photons trigger the image intensifier and Timepix4 readout. The scintillator type and thickness affect neutron absorption, light yield, and decay behavior. Finally, the pulsed time structure of the beamline concentrates neutron arrivals in certain TOF bins, making the maximum instantaneous pixel-hit rate substantially higher than the time-averaged value.

During these experiments, CSNS operated at a source power of 170 kW. To estimate the readout load at 500 kW, we assumed that the pixel-hit rate scales linearly with source power while the beamline and detector conditions remain unchanged. Based on the measured maximum instantaneous pixel-hit rate per pixel of 1662.9 hits/pixel/s under the 0.14\ensuremath{\times} magnification configuration, a full-matrix extrapolation in which all pixels of the Timepix4 matrix operate at this level simultaneously gives a chip-level instantaneous pixel-hit rate of 381.4 Mhits/s. Scaling this value to a 500 kW source power gives an estimated rate of 1.12 Ghits/s. Under these assumptions, the estimated load is below the approximately 2.5 Ghits/s chip-level Timepix4 specification \cite{ref38}, and is well above the 80 Mhits/s readout limit of Timepix3 \cite{ref32}. This is a conditional extrapolation rather than a measured whole-chip rate, and it does not include image-intensifier saturation, optical-event overlap, reconstruction losses, packet overhead, or other system-level bottlenecks. This comparison supports the choice of higher-throughput readout, while the measured pulsed-beam load provides a quantitative basis for assessing readout requirements in scintillator-based event-mode neutron imaging.

\subsection{Imaging demonstrations}\label{sec:imaging-demonstrations}

\subsubsection{Conventional radiography}\label{sec:conventional-radiography}

\begin{figure}[!htb]
\centering
\includegraphics[width=\columnwidth]{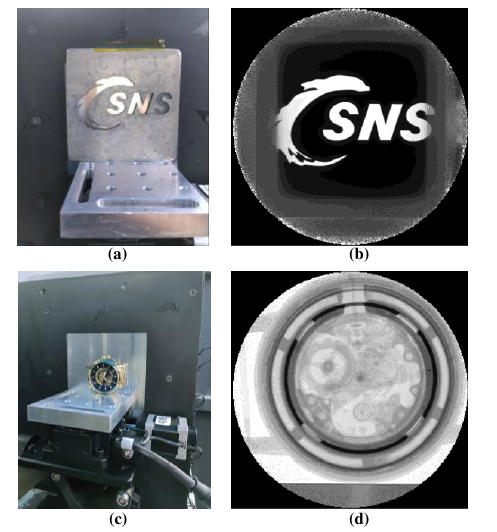}
\caption{Photographs and corresponding neutron radiographs: (a) photograph of the Cd CSNS logo sample; (b) neutron radiograph of the logo; (c) photograph of the mechanical wristwatch; (d) neutron radiograph of the wristwatch.}
\label{fig:11}
\end{figure}

Although the event-mode data preserve timing information, they can also be accumulated without wavelength subdivision to form conventional neutron radiographs. Under the 0.14\ensuremath{\times} magnification configuration, a cadmium plate with a cut-out CSNS logo was imaged to demonstrate large-FOV radiography. Under the 0.35\ensuremath{\times} magnification configuration, neutron radiography of a mechanical wristwatch was performed to demonstrate fine structural visualization. The Cd logo sample, shown in Fig.~\ref{fig:11}(a), has dimensions of 10 cm \ensuremath{\times} 10 cm. As shown in Fig.~\ref{fig:11}(b), the overall logo shape and internal structural features are clearly resolved. Some edge regions are less clearly resolved, which may reflect limited beam coverage and contributions from scattered neutrons. Figs.~\ref{fig:11}(c) and \ref{fig:11}(d) show the photograph and neutron radiograph of the mechanical wristwatch. Metallic components such as the watch case and gears exhibit high neutron transmission, whereas hydrogen-rich organic components such as rubber gaskets show stronger attenuation and appear darker. By exploiting the high penetration of neutrons in metals and their strong interaction with hydrogen-rich materials, this imaging approach enables nondestructive inspection of organic components embedded within metallic structures.

\subsubsection{Energy-selective imaging}\label{sec:energy-selective-imaging}

Energy-selective imaging uses reconstructed neutron-event arrival times to select wavelength bands and form the corresponding transmission images, whereas conventional radiography forms a broadband transmission image. Under the 0.14\ensuremath{\times} magnification configuration, simultaneous neutron radiography was performed on seven metallic samples: cobalt (Co), molybdenum (Mo), tungsten (W), iron (Fe), nickel (Ni), copper (Cu), and brass. Each sample had a cylindrical geometry with a diameter of 6 mm and a length of 100 mm and was arranged close to the detector, with its cylinder axis perpendicular to the neutron beam direction. Transmission images were obtained as pixelwise ratios of reconstructed neutron-event count rates in the sample measurement to those in the open-beam measurement. Fig.~\ref{fig:12}(a) shows the wavelength-selected transmission images, where brighter regions correspond to higher transmission ratios. In the broadband image labeled 1--5.5 \AA{}, cobalt exhibits the lowest transmission and molybdenum the highest, while tungsten and nickel show comparable overall transmission. Their contrast changes with the selected wavelength band: tungsten shows higher transmission than nickel in the 1--2 \AA{} band, whereas nickel exhibits higher transmission than tungsten in the 4--5 \AA{} band. This contrast reversal demonstrates the wavelength-dependent attenuation information provided by the event-mode data.

\begin{widefigurehere}
\centering
\includegraphics[width=129mm]{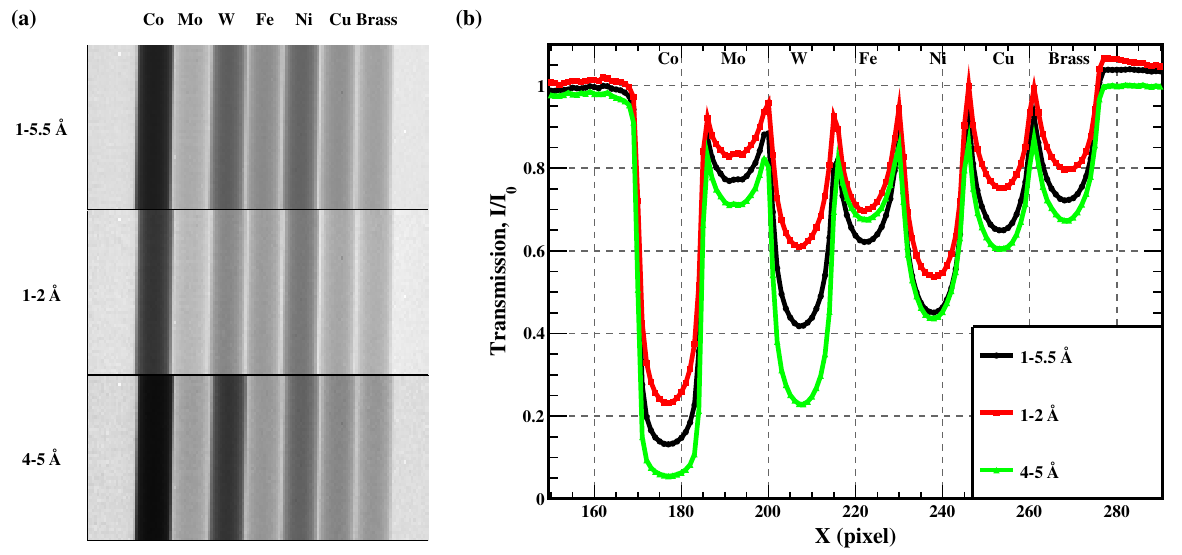}
\caption{Energy-selective neutron imaging of seven metallic samples: (a) wavelength-selected transmission images in three wavelength bands; (b) corresponding X-direction transmission profiles.}
\label{fig:12}
\end{widefigurehere}

To quantitatively evaluate the wavelength-dependent transmission variations, Fig.~\ref{fig:12}(b) shows the X-direction transmission profiles of the seven samples. The profiles were obtained by averaging the pixelwise transmission ratios over Y = 230--280. The ratios outside the sample projections are generally close to unity but show wavelength-dependent deviations. These deviations may be associated with additional counts from sample-scattered neutrons on the right \cite{raventos2018,pfretzschner2020} and absorption of obliquely incident or scattered background neutrons by the neighboring Co sample on the left. Within the sample projections, tungsten and nickel exhibit similar minimum profile values of approximately 41.8\% and 45.1\% in the broadband image labeled 1--5.5 \AA{}. Wavelength selection increases the difference between their transmission values and reverses their ordering: in the 1--2 \AA{} band, the minimum profile values are 61.0\% and 53.7\%, whereas in the 4--5 \AA{} band they are 22.9\% and 43.7\%, respectively. These results demonstrate that wavelength selection based on reconstructed neutron-event arrival times can enhance the contrast between materials with similar broadband neutron transmission.

\subsubsection{Bragg-edge imaging}\label{sec:bragg-edge-imaging}

Neutron Bragg-edge imaging provides crystallographic information and also offers a way to estimate the effective relative wavelength resolution under the measurement conditions. Using an austenitic steel sample with a face-centered cubic (FCC) structure, Bragg-edge imaging was performed under the 0.35\ensuremath{\times} magnification configuration. The transmission spectrum of the sample is shown in Fig.~\ref{fig:13}(a). At neutron wavelengths of 2.17 \AA{}, 2.54 \AA{}, 3.6 \AA{}, and 4.15 \AA{}, the corresponding Bragg edges were identified as FCC (311), FCC (220), FCC (200), and FCC (111), respectively. These positions are consistent with literature values \cite{ref45,ref46,ref47}, demonstrating that the Timepix4-readout event-mode imaging approach can perform Bragg-edge neutron imaging. 

\begin{columnfigurehere}
\centering
\includegraphics[width=\columnwidth]{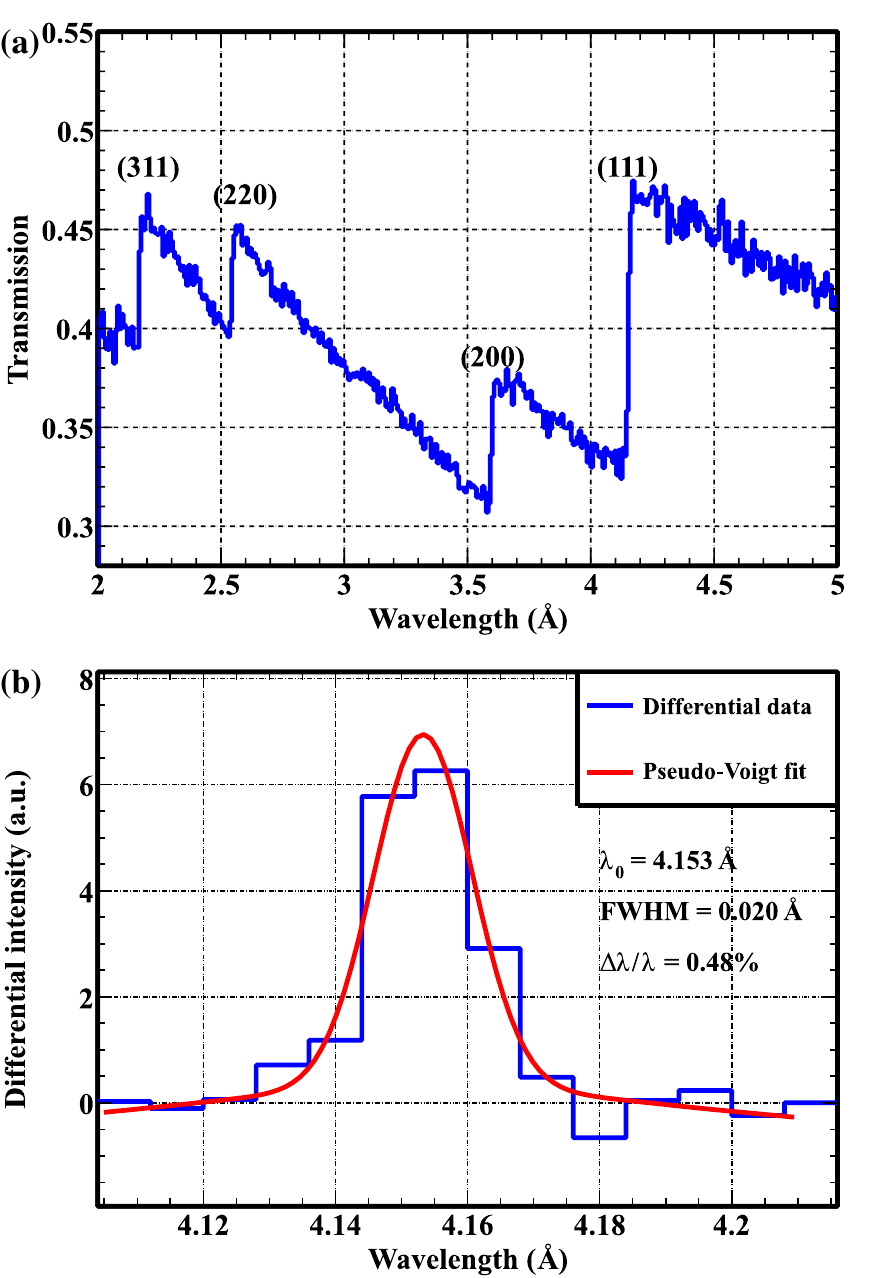}
\caption{Bragg-edge imaging results of an austenitic steel sample: (a) transmission spectrum; (b) derivative of the FCC (111) Bragg edge with a pseudo-Voigt fit.}
\label{fig:13}
\end{columnfigurehere}

The FCC (111) Bragg edge was differentiated, and the resulting curve is shown in Fig.~\ref{fig:13}(b). Here, FWHM denotes the full width at half maximum. Fitting of the differentiated data using a pseudo-Voigt function \cite{ref45,ref48} yielded an FWHM of \ensuremath{\Delta}\ensuremath{\lambda} = 0.020 \AA{}, corresponding to a relative width of \ensuremath{\Delta}\ensuremath{\lambda}/\ensuremath{\lambda} = 0.48\% at the 4.15 \AA{} edge. This value provides a Bragg-edge-based estimate of the effective relative wavelength resolution under the present measurement conditions. Together with the identified FCC Bragg edges, this result demonstrates the use of reconstructed neutron events to obtain crystallographic features from wavelength-dependent transmission measurements.

\section{Conclusion and outlook}\label{sec:conclusion-and-outlook}

Scintillator-based event-mode neutron imaging was implemented using a self-developed Timepix4-based optical readout system and investigated through image-formation analysis and beam experiments at CSNS. The analysis distinguishes the spatial extent of scintillation responses from event-localization error, providing a basis for interpreting the spatial imaging benefit of event reconstruction. Under the same 0.35\ensuremath{\times} optical configuration, reconstruction improved the spatial resolution from 164.2 to 91.9 \ensuremath{\mu}m at 10\% MTF. Following event reconstruction and rejection of single-photon reconstructed events, the mean peak-normalized response in the 6--8 \AA{} out-of-band interval was 85.4\% lower than in the raw-pixel wavelength-count spectrum, indicating reduced afterglow-related spectral background. Sample measurements demonstrated conventional radiography, energy-selective imaging, and Bragg-edge imaging using reconstructed neutron-event positions and arrival times. An effective relative wavelength resolution of 0.48\% was estimated from the FCC (111) Bragg edge under the present experimental conditions. Together, the image and spectrum comparisons show that measurable imaging-quality gains can be obtained through event reconstruction and rejection of single-photon reconstructed events without changing the scintillator screen or optical configuration. Under the pulsed-beam conditions tested in this study, the maximum instantaneous pixel-hit rate per pixel was approximately four times its time-averaged value. This measured peak-to-average ratio illustrates the importance of accounting for the temporal distribution of pixel hits when assessing readout requirements at higher neutron flux. Further development should improve the match between the image-intensifier active area and the Timepix4 sensitive area, optimize scintillator selection and optical coupling, and evaluate system performance under higher-flux conditions relevant to CSNS Phase II.

\section*{Acknowledgments}

This work was supported by the National Key Research and Development Program of China (No. 2023YFC2206500), the National Natural Science Foundation of China (Nos. 12305348, 12227810 and 12527806), the Guangdong Basic and Applied Basic Research Foundation (Nos. 2025A1515011320, 2024B1515120006 and 2023A1515110444), the Guangdong Major Project of Basic and Applied Basic Research (Grant No. 2023B0303000003) and the Guangdong Provincial Key Laboratory of Advanced Particle Detection Technology (No. 2024B1212010005).

\FloatBarrier
% Preserve the normal declaration-heading separation after acknowledgments.
\vspace*{1.5\baselineskip}
\section*{Author Contributions}

All authors contributed to the study conception and design. Material preparation, data collection and analysis were performed by Wenqin Yang, Junan Geng, and Xingfen Jiang. The first draft of the manuscript was written by Wenqin Yang, and all authors commented on previous versions of the manuscript. All authors read and approved the final manuscript.

\section*{Data Availability Statement}

The data supporting the findings of this study are openly available in Science Data Bank at \url{https://doi.org/10.57760/sciencedb.j00186.0013b} and \url{https://cstr.cn/31253.11.sciencedb.j00186.0013b}.

\section*{Conflict of Interest}

The authors declare that they have no conflict of interest.

\section*{Generative AI Statement}

\textit{Declaration of Generative AI Use:} During the preparation of this work, the authors used ChatGPT (OpenAI) and DeepSeek for language editing and grammar checking only. These tools were used to improve the readability and clarity of the text. After using these tools, the authors reviewed and edited the content as needed and take full responsibility for the content of the manuscript. No AI tool was used to generate scientific data, perform analysis, or formulate conclusions.

\end{document}